\documentclass[aps,prl,twocolumn,floatfix,showpacs,amssymb]{revtex4-1}
\usepackage[utf8]{inputenc}
\usepackage[english]{babel}
\usepackage{amsmath}
\usepackage{amsfonts}
\usepackage{amssymb}
\usepackage{graphicx}
\usepackage{float}
\usepackage{color}
\usepackage{braket}
\usepackage{mciteplus}
\def\url#1{}

\begin{document}

\title{Multiband Mechanism for the Sign Reversal of Coulomb Drag Observed in Double Bilayer Graphene Heterostructures}

\author{M. Zarenia$^{1,2}$, A. R. Hamilton$^3$, F. M. Peeters$^1$ and D. Neilson$^{1,4}$}
\affiliation{$^1$Department of Physics, University of Antwerp, Groenenborgerlaan 171,  2020 Antwerp, Belgium\\
$^2$Department of Physics and Astronomy, University of Missouri, Columbia, Missouri 65211, USA\\
$^3$ARC Centre of Excellence for Future Low Energy Electronics Technologies,
School of Physics, The University of New South Wales, Sydney NSW 2052, Australia\\
$^4$Physics Division, Science \& Technology School, University of Camerino, 62032 Camerino, Italy}

\begin{abstract}

Coupled 2D sheets of electrons and holes are predicted to support novel quantum phases.
Two  experiments of Coulomb drag in electron-hole  (e-h)
double  bilayer graphene (DBLG) have reported
an unexplained and puzzling sign reversal of the drag signal.
However, we show that this effect is due to
the multiband character of DBLG.
Our multiband Fermi liquid theory
produces excellent agreement and
captures the key features of the experimental drag resistance  for all temperatures.
This demonstrates the importance of multiband effects in DBLG:
they  have a strong effect not only on superfluidity, but also on the drag.

\end{abstract}
\maketitle

\begin{figure}[t]
\includegraphics[width=8.5cm]{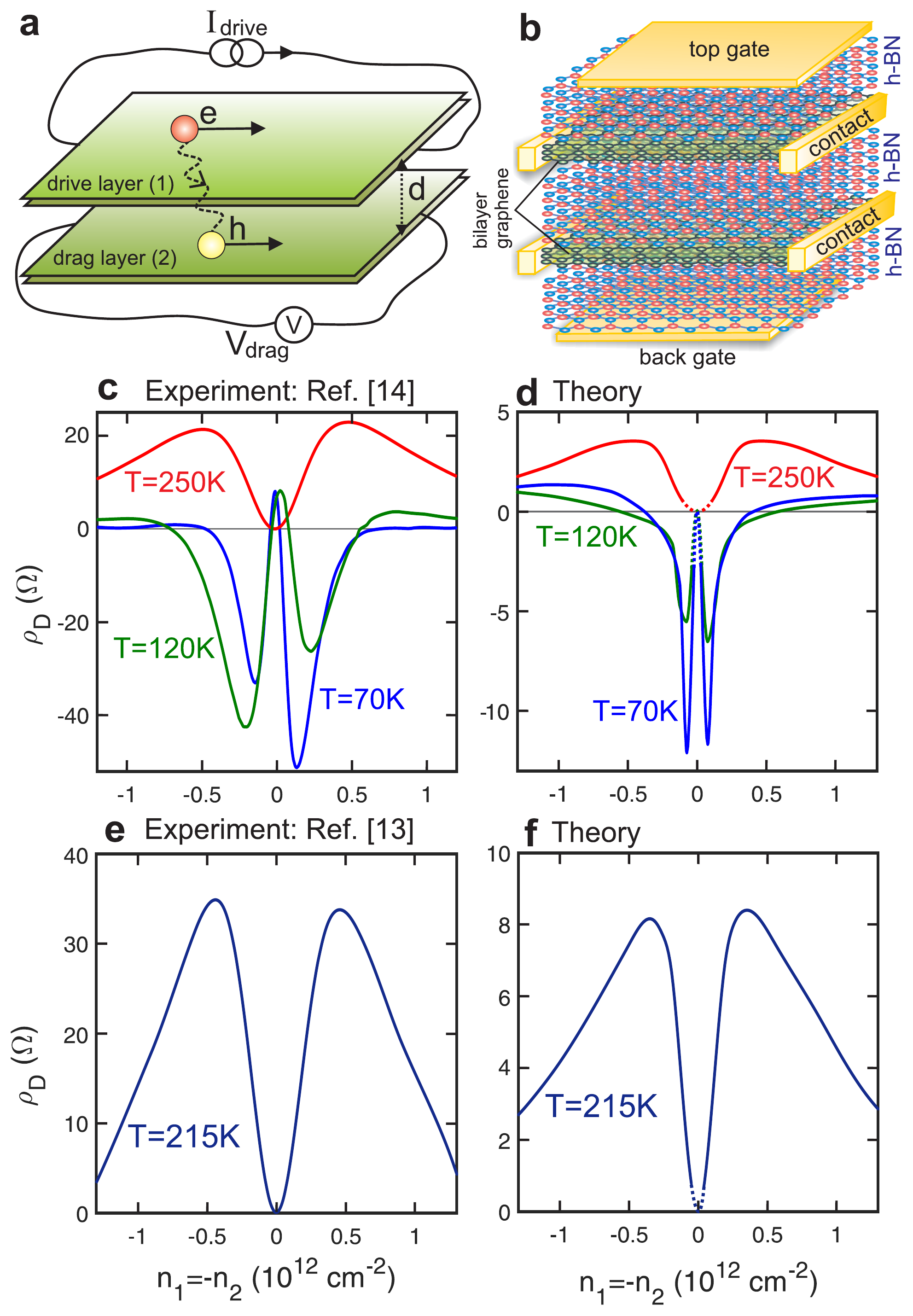}
\caption{
(a) Schematic representation of the e-h drag measurement.
(b) Schematic of the device.
(c) Experimental drag measurements from  Ref.\ \citenum{li_negativeDrag}.
(d) Our calculated drag for sample in Ref.\ \citenum{li_negativeDrag}.
(e) Experimental data from Sample A of Ref.\  \citenum{lee_giantDrag}.
(f) Our calculated drag for sample A in Ref.\ \citenum{lee_giantDrag}.
}
\label{e-hDrag_th&exp}
\end{figure}

Electron-hole (e-h) double-sheet van der Waals heterostructures
are attracting great interest because they are predicted to support novel quantum phases.
These phases include superfluidity, coupled Wigner crystals,
and charge density waves.\cite{lozovik_superc,Eisenstein2004,Swierkowski1991,DePalo2002,perali_dbg,Zarenia2017}
Novel quantum phases are  a major motivator
for  experimental studies of Coulomb drag
in coupled e-h sheets in:
GaAs double quantum wells\cite{croxall_anomalous,seamons_GaAs},
graphene double monolayers\cite{kim_coulombdrag,gorbachev_strong},
hybrid graphene-GaAs systems\cite{Gamucci2014,Simonet2017},
and graphene double bilayers (DBLG)\cite{lee_giantDrag,li_negativeDrag}.
In a drag measurement,
the measured quantity is the transresistivity $\rho_D$,
the ratio of the  generated voltage in the open-circuit drag sheet (2)
to the current density in the drive sheet (1)
(Fig.\ \ref{e-hDrag_th&exp}(a)).
In conventional momentum drag, the dragged hole is expected to travel
in the same direction as the drive electron,
corresponding to a positive drag resistivity, $\rho_D>0$.

Deviations of $\rho_D$ from
a standard Fermi-liquid $T^2$ temperature dependence,  are commonly accepted as
evidence of correlations\cite{Swierkowski1995}
or the existence of exotic many-body phases\cite{Vignale1996,Efimkin2016}.
Two recent experimental studies of e-h drag in DBLG\cite{lee_giantDrag,li_negativeDrag}
reporting large anomalous behavior in the $\rho_D$
have therefore attracted a lot of attention and discussion.
Even more puzzling,
Lee {\it et al.}\cite{lee_giantDrag}  reported  $\rho_D$ that reverse sign
as the carrier density was decreased at low $T$.  The magnitudes of the $\rho_D$
increase dramatically with decreasing $T$,
eventually becoming comparable to the sheet resistivity.
Independently, Li {\it et al.}\cite{li_negativeDrag} also reported $\rho_D$ that reversed sign
as a function of the  density
for $T$ as high as $160$ K.
These are completely unexpected results, with the reversal of sign in the $\rho_D$ a major conundrum.
The $\rho_D$ in double GaAs quantum wells\cite{Gramila1991,Swierkowski1995}
and double  graphene monolayers\cite{gorbachev_strong},
show no sign reversal and
their drag mechanisms are substantially understood.\cite{Narozhny2016,hwang_coulomb}
However there exists no explanation for the recent DBLG results.

We show that these striking effects
can be explained in detail by the multiband character of bilayer graphene (BLG) taken together
with the increase in the carrier effective masses ($m^{\ast}$) at low  densities and
$T$.
Our theory of a linear screened Fermi-liquid in multiband BLG
produces excellent agreement with the observed structure in the $\rho_D$,
capturing the key features of the recent experiments over the full  range of $T$
and highlighting the dramatic consequences of multiband transport on the drag.

%
Figure \ref{e-hDrag_th&exp}(b) schematically depicts
electron and hole BLG sheets, ($1$) and ($2$),
encapsulated in insulating hexagonal Boron Nitride (hBN)
to prevent tunneling between the sheets and e-h recombination.
Metal gates control the
sheet densities $n_1$ and $n_2$.
Unlike monolayer graphene,
BLG has  single-particle energy bands with quadratic dispersion at low-energies,
and its conduction band (CB) and valence band (VB) are separated by a small variable bandgap
induced by the perpendicular electric fields from the
gates.
In contrast, monolayer graphene has linearly dispersing bands with no bandgap, while
GaAs has quadratic bands with a fixed large bandgap $\gg k_B T$.
We find in DBLG that there is significant multiband transport  in both the CB and VB
of each sheet which must be included in calculations.
An understanding of the nature of the multiband mechanisms lying behind these surprising experimental results
paves the way
for future drag experiments.
Multiband effects have already been shown to be important
for e-h superfluidity in coupled 2D materials.\cite{Conti2017}

Figures \ref{e-hDrag_th&exp}(c)-\ref{e-hDrag_th&exp}(f)  compare our $\rho_D$ at fixed $T$ as a function of
matched  densities $n_1=-n_2$
with the data of Refs.\ \citenum{lee_giantDrag},\citenum{li_negativeDrag}.
Figure \ref{e-hDrag_th&exp}(c) shows data from  Ref.\ \citenum{li_negativeDrag} of the drag.
For all $T$, the system displays conventional positive drag at high densities, but
unexpected sharp negative peaks appear as the density is reduced  at lower $T$.

Figure \ref{e-hDrag_th&exp}(d) shows our calculated $\rho_D$
for the  experimental parameters in Ref.\ \citenum{li_negativeDrag}: hBN barrier thickness $d=5.2$ nm;
the transverse electric fields
at the dual neutrality point (DNP) due to unintentional doping  are not precisely specified  and
we take reasonable values, $(E^0_1,E^0_2)=(0.05,0)$ meV.nm$^{-1}$.
Our results in Fig.\ \ref{e-hDrag_th&exp}(d) capture the
key structure of the unusual positive and negative peaks in the experimental $\rho_D$
at fixed $T$, as well as the
change of sign of the $\rho_D$ when $T$ is increased.
The asymmetries in the negative peaks in Figs.\ \ref{e-hDrag_th&exp}(c) and \ref{e-hDrag_th&exp}(d) are reversed.
We will see later (Fig.\ \ref{e-hDrag}(b)) that, unlike the positions of the peaks, their asymmetries are very sensitive to  $E_1^0$ and $E_2^0$,
and small changes can reverse the asymmetry.  Lacking precise experimental values for $E_1^0$ and $E_2^0$,
we chose not to attempt to reproduce the asymmetries through small adjustments of $E_1^0$ and $E_2^0$.

Consistent with experiment, we obtain positive e-h drag for
$T = 215$ K and $250$ K, but strong negative drag at small densities for $T=70$ K and $120$ K.
A discrepancy is  that our theory does not reproduce
the small positive peak in $\rho_D$ observed at the DNP for $T=70$ K and $120$ K.
This peak is known to be associated with  energy drag from
e-h puddle density fluctuations not considered by our theory.
Our $\rho_D$ must  vanish at the DNP, and very close to the DNP we
extrapolate $\rho_D$  to zero (dotted lines).
Our theory also reproduces the key features of drag measured in different samples in another laboratory.
Figure \ref{e-hDrag_th&exp}(e) shows data from Sample A of Ref.\  \citenum{lee_giantDrag}, and
Fig.\ \ref{e-hDrag_th&exp}(f)
the calculated drag using the experimental  parameters  $d=3$ nm and
$(E^0_1,E^0_2)=(0.1,-0.04)$ meV.nm$^{-1}$.


We use a well-established procedure for determining
the drag transresistivity $\rho_D$ in e-h DBLG
at fixed $T$ as a function of equal
densities, $n_1=-n_2$.
We use the Random Phase Approximation (RPA) linear-response screening theory
for Fermi liquids in the clean limit.
The consistent underestimate of the magnitude of $\rho_D$  in our model
may be due to many-body correlations,
which are expected to increase the magnitude of $\rho_D$
but not to strongly modify its overall shape.\cite{Swierkowski1995}
New in our approach is the inclusion of the induced variable
bandgaps
between the conduction and valence bands of
the bilayers,
the link  between the variable
bandgaps
 and the carrier densities,
and the increase in $m^\ast$ near the DNP at low $T$.

$\rho_D$ is related to
the  drag conductivity $\sigma_D$ by,\cite{Rojo1999,Narozhny2016}
\begin{equation}\label{rhoD}
\rho_D=-\sigma_D/[\sigma_1\sigma_2-\sigma_D^2] \simeq -\sigma_D/\sigma_1\sigma_2\ .
\end{equation}
$\sigma_{i}=n_{i}e^2\tau_{i}/m^{\ast}$ is the longitudinal conductivity for sheet ($i$)
with momentum independent scattering time, $\tau_1\approx\tau_2 \equiv \tau$.
Charge impurity scattering and short-range impurity scattering both lead to
transport scattering
times
independent of momentum in BLG\cite{hwang_coulomb}.
When there are transverse electric fields at the DNP from unintentional doping,
$m^{\ast}$ is expected to be enhanced at low densities\cite{McCann2006},
as recently observed experimentally\cite{lee_chemical,lee_giantDrag}.

\begin{figure*}[t]
\includegraphics[scale=0.42]{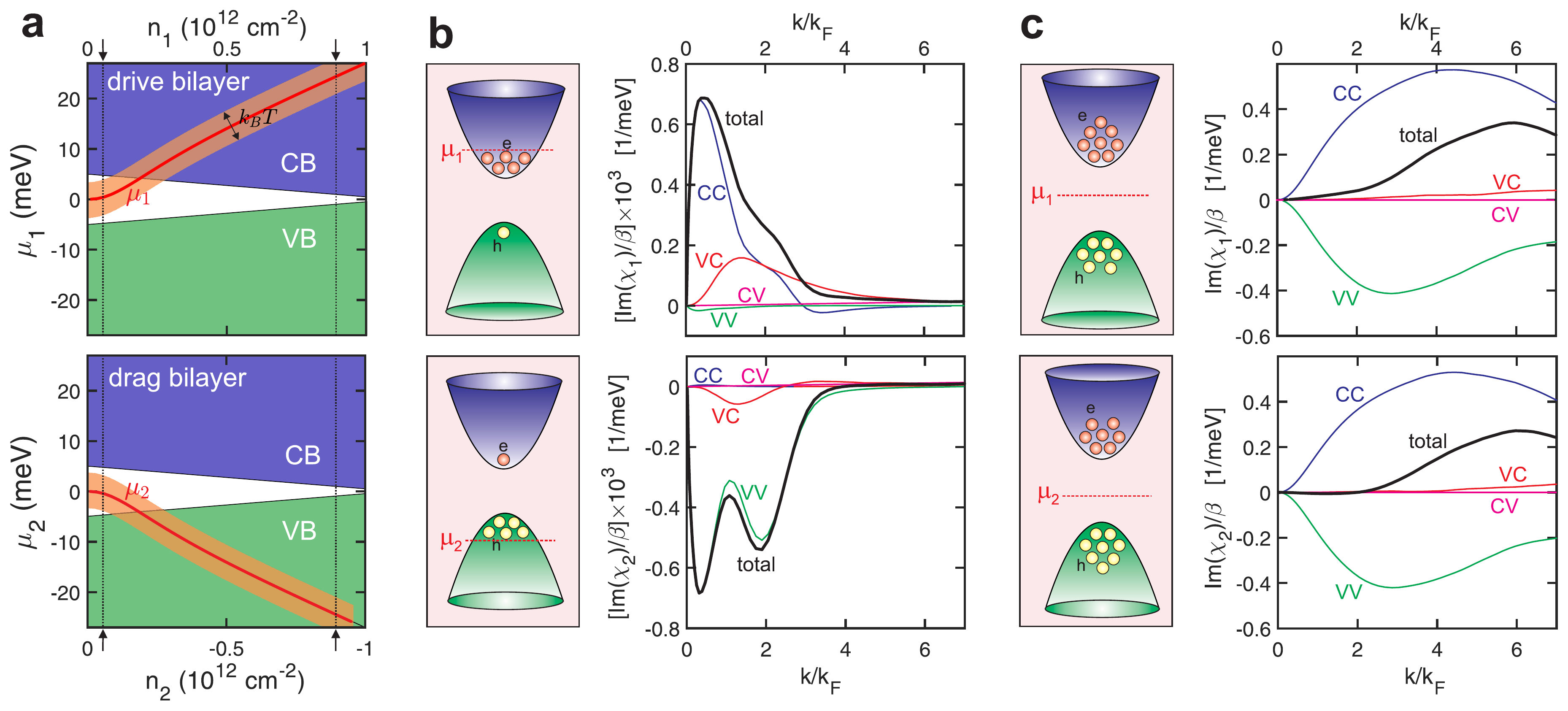}
\caption{
Physics behind the change of sign in $\rho_D$. We set $T=70$ K.
(a) Evolution of  the bands with  density in the
$e$- and $h$-bilayers ($1$) and ($2$).
The orange areas represent  thermal excitations.
(b) Left panels: carrier  populations for a higher
density $n_1=-n_2=0.8\times10^{12}$ cm$^{-2}$.
Right panels: contributions in Eq.\ (\ref{eqX}),
$\mathfrak{Im}\chi^{\gamma\gamma'}_1(\boldsymbol{q},\omega)$ and
$\mathfrak{Im}\chi^{\gamma\gamma'}_2(\boldsymbol{q},\omega)$ (colored curves),
for  $\hbar\omega=1$ {meV}.
$\beta= e \tau \hbar k_F^3/(2m^{\ast})$.
Dominant contributions $\mathfrak{Im}\chi^{CC}_1$ and $\mathfrak{Im}\chi^{VV}_2$
have opposite signs so $\rho_D \varpropto -(\mathfrak{Im}\chi_1 \mathfrak{Im}\chi_2) >0$.
(c) Left panels: the comparable CB and VB populations for lower density
$n_1=-n_2=0.05\times10^{12}$ cm$^{-2}$.
Right panels:  contributions to
$\mathfrak{Im}\chi^{\gamma\gamma'}_1(\boldsymbol{q},\omega)$ and $\mathfrak{Im}\chi^{\gamma\gamma'}_2(\boldsymbol{q},\omega)$.
The terms for each $(\gamma\gamma')$ are  similar for both sheets,
so $\rho_D \varpropto -(\mathfrak{Im}\chi_1 \mathfrak{Im}\chi_2) <0$.
}
\label{populations}
\end{figure*}

$\sigma_D$ is a convolution of the density fluctuations within the  sheets
represented by $\mathfrak{Im}\chi_i(\boldsymbol{q},\omega),i=1,2$,
the imaginary part of the nonlinear susceptibility of each sheet,
\begin{equation}
\label{sigmaD}
\sigma_{D} =
\frac{\hbar}{4\pi k_BT}\!\int\! d^2\boldsymbol{q} d\omega\frac{|V_{12}(\boldsymbol{q},\omega)|^2
\mathfrak{Im}\chi_1(\boldsymbol{q},\omega)
\mathfrak{Im}\chi_2(\boldsymbol{q},\omega)}{\sinh^2{\hbar\omega/2k_BT}}
\end{equation}
\begin{eqnarray}
\label{eqX}
\mathfrak{Im}\chi_{i}(\boldsymbol{q},\omega) =
e \tau\sum_{\gamma\gamma^\prime} \mathfrak{Im} \left\{
\int d^2\boldsymbol{k} \right.
\ \ \ \ \ \ \ \ \ \ \ \ \ \ \ \ \ \ \ \ \ \ \
\ \nonumber\\
\left. \frac{F_{\boldsymbol{k},\boldsymbol{q}}^{\gamma\gamma^\prime}
\left(f_{\epsilon_{\boldsymbol{k},\gamma}}^{i}-f_{\epsilon_{\boldsymbol{k}+\boldsymbol{q},\gamma'}}^{i}\right)
\left(v_{\boldsymbol{k},\gamma}^{x,i}-v_{\boldsymbol{k}+\boldsymbol{q},\gamma^\prime}^{x,i}\right)}
{\varepsilon_{\boldsymbol{k},\gamma}^{i}-\varepsilon_{\boldsymbol{k}+\boldsymbol{q},\gamma^\prime}^{i}
+\hbar\omega+i0^{+}}\right\}.
\end{eqnarray}
$\gamma=C$,$V$ labels the CB and VB.
The form factors $F_{\boldsymbol{k},\boldsymbol{q}}^{\gamma\gamma^\prime}$
come from the overlap of the
wave-functions in  the gapped BLG.\cite{Wang2010}
In band $\gamma$ of sheet ($i$),
$f_{\varepsilon_{\boldsymbol{k},\gamma}}^{i}=1/[e^{{(\epsilon_{\boldsymbol{k},\gamma}-\mu_i)}/k_BT}+1]$
is the Fermi-Dirac  function with chemical potential $\mu_i$,
$v_{\boldsymbol{k},\gamma}^{x,i}$ is the $x$-component of the velocity, and
$\varepsilon_{\boldsymbol{k},\gamma}^{i}$ is the single-particle energy,
obtained using the four-band Hamiltonian for biased BLG
subbands\cite{McCann2013} with variable bandgap $E_g^{i} $.
For
the small gaps we work with, $E_g^{i} < 20$ meV,
and for the low densities, $n=p< 10^{12}$ cm$^{-2}$,
our results are not changed significantly
if we use the quadratic energy dispersion
obtained with the corresponding two-band Hamiltonian.

In the  $|\mu|\rightarrow 0$ limit, the analytic solutions for $\rho_D$ in a clean system
become  numerically tedious to solve, and since  $\rho_D\rightarrow 0$ for $|\mu|/k_B T \ll 1$,
we  extrapolate it to zero in this limit.
This extrapolation may not even be relevant to experiments, since it will be
masked by the peak in $\rho_D$ due to disorder-induced e-h puddle density fluctuations\cite{Song2012}
that are absent in the clean system.

The bandgap $E_g^{i}$ in  sheet ($i$) depends on
the transverse electric fields $E_i$\cite{Castro2007,zhang_bandgap}
from the metallic gates.
Thus
$E_g^{i}$ changes slightly with the carrier density $n_i$.
Through the doping of each bilayer, $E_i$
can be described as (see
supplementary information,
Ref.\ \citenum{lee_giantDrag}),
\begin{eqnarray}\label{eqE1E2}
E_i &=& \delta_{i2}(en_2/2\epsilon_0) + en_1 /\epsilon_0 + E_i^0\ .
\end{eqnarray}
According to the experimental data from Refs.\ \citenum{zhang_bandgap,lee_chemical},
the induced bandgap $E_g^i\approx \alpha E_i$, with  $\alpha\sim 0.1e$ C.nm for weak $E_i$, and
$E_g^i$ and $E_i$  in
meV and mV.nm$^{-1}$.
Typically in the samples
in Refs.\ \citenum{lee_giantDrag,li_negativeDrag},  $E_g^i\lesssim20$ meV.

At low $T$ when the system is degenerate, we take a typical
value for the gap at the DNP of $E_g^i=20$ meV to calculate
the enhancement of  $m^{\ast}$ at small densities,
using the density-dependent expression for $m^{\ast}$ in Ref.\ \citenum{McCannMass}.
At  room temperature,  the number of thermally excited carriers will be large enough
to suppress this enhancement.
Hence we interpolate $m^{\ast}$ from the density-dependent values for low $T$
to the unrenormalized, constant value $m^{\ast} = 0.04 m_e$\cite{zou_effectivem} at room temperature.

The dynamically screened Coulomb interaction between sheets ($i$) and ($j$) within the RPA is
$V_{ij}(\boldsymbol{q},\omega) = (-1)^{i-j}v(q)\exp[-qd(1-\delta_{ij})]/\det|\epsilon(\boldsymbol{q},\omega)|$,
where $v(q)=2\pi e^2/\kappa q$ is the bare Coulomb interaction and
$\epsilon_{ij}(\boldsymbol{q},\omega)=\delta_{ij}+V_{ij}(\boldsymbol{q},\omega)\Pi_i(\boldsymbol{q},\omega)$,
with $\Pi_i(\boldsymbol{q},\omega)$  the RPA polarization function for sheet ($i$).\cite{Wang2010}
At low electric fields, $E_i\lesssim 10$ meV.nm$^{-1}$,
%
%
the $\kappa$ for an isolated BLG encapsulated in hBN layers increases from $2$ for few layers to $4$ for many layers.\cite{Kumar2016}
For our DBLG encapsulated in hBN, we expect $2\lesssim \kappa \lesssim 4$.
Across this range of $\kappa$, we find the shape of the drag peaks in $\rho_D$ does not significantly vary,
but their height increases with increasing $\kappa$ (the strength of
screening in $V_{ij}(\boldsymbol{q},\omega)$ depends on $\kappa$).
For our results we have set $\kappa = 4$.


We numerically solve Eqs.\ (\ref{rhoD}) to (\ref{eqX})  for $\rho_D$.  We fix
the chemical potential $\mu_i$ for sheet ($i$) using,
\begin{equation}\label{eqMu}
n_i =
4\int d^2\boldsymbol{k}
\left[f_{\varepsilon_{\boldsymbol{k},\gamma=1}}^{i}+\big(f_{\varepsilon_{\boldsymbol{k},\gamma=-1}}^{i}-1\big)\right]\ .
\end{equation}

To understand the origins of the drag response  structure, in Fig.\ \ref{populations} we examine
the four terms in the summation in Eq.\ (\ref{eqX})
over the CB and VB,  $\gamma=C,V$ and $\gamma'=C,V$.
For Fig.\ \ref{populations} we set $T=70$ K, comparable to the bandgaps $E_g^1=E_g^2$ that
correspond to  $E^0_1=-E^0_2=0.05$ meV.nm$^{-1}$.
Figure \ref{populations}(a)  shows schematically the evolution
of the CB and VB structure with density in the
e-doped drive bilayer ($1$) and
h-doped  drag bilayer ($2$).
Unlike  conventional semiconductors, the bandgaps are small and  depend on density.
At higher densities, the chemical potentials $\mu_i$ (red lines) are in the CB and VB,
leading  to  asymmetric band populations,
schematically represented
in the left panels of Fig.\ \ref{populations}(b).
The large contributions to the summations in Eq.\ (\ref{eqX}) are
the intraband terms from the CB in bilayer (1), $\mathfrak{Im}\chi^{CC}_1$
(blue curve in right panels of Fig.\ \ref{populations}(b)),
and
the VB in bilayer (2), $\mathfrak{Im}\chi^{VV}_2$ (green curve).
Since $\mathfrak{Im}\chi^{CC}_1$ and $\mathfrak{Im}\chi^{VV}_2$ have opposite signs,
the total $\mathfrak{Im}\chi_1(\boldsymbol{q},\omega)$ and $\mathfrak{Im}\chi_2(\boldsymbol{q},\omega)$
(black curves)
also have opposite signs.   The e-h drag $\rho_D \varpropto -(\mathfrak{Im}\chi_1 \mathfrak{Im}\chi_2)$
is then everywhere positive, as in a conventional semiconductor.

Figure  \ref{populations}(c) is at a smaller density where the  $\mu_i$ lie near the midgaps.
Thermal fluctuations  ensure  there are significant
carrier populations in the CB and VB (left panels Fig.\  \ref{populations}(c)),
which makes each term in Eq.\  (\ref{eqX}) similar for the two sheets,
$\mathfrak{Im}\chi^{\gamma\gamma'}_1\approx \mathfrak{Im}\chi^{\gamma\gamma'}_2$
(compare the curves in right panels of Fig.\  \ref{populations}(c)).
Since $\mathfrak{Im}\chi_1(\boldsymbol{q},\omega)$ and $\mathfrak{Im}\chi_2(\boldsymbol{q},\omega)$
(black curves)  are similar and have the same sign,
$\rho_D \varpropto -(\mathfrak{Im}\chi_1 \mathfrak{Im}\chi_2)$ is negative.
This is the key mechanism behind the sign change in $\rho_D$.
A change of sign occurs with increasing density when $\mu$ moves from the midgap
to one of the bands, and it thus represents
a crossover from multiband to single-band physics.

At higher $T$ (not shown), the $\mu_i$ evolve with increasing density
similarly to Fig.\ \ref{populations}(a).
However this evolution occurs much faster because of the smaller $m^\ast$ at higher $T$.
For $T\gtrsim 200$ K, the negative peaks in $\rho_D$
are confined to such an extremely
narrow band of densities around the DNP that they would be practically unobservable due to disorder, and
the $\rho_D$ is essentially everywhere positive.
We note that the renormalization of $m^\ast$ at low $T$ is essential for the negative peaks;
if $m^\ast$ is not renormalized, the negative drag
in Fig.\ \ref{e-hDrag_th&exp}d for $T=70$ K and $120$ K would similarly be unobservable,
being narrowly confined around the DNP.

\begin{figure}[t]
\includegraphics[width=8.5cm]{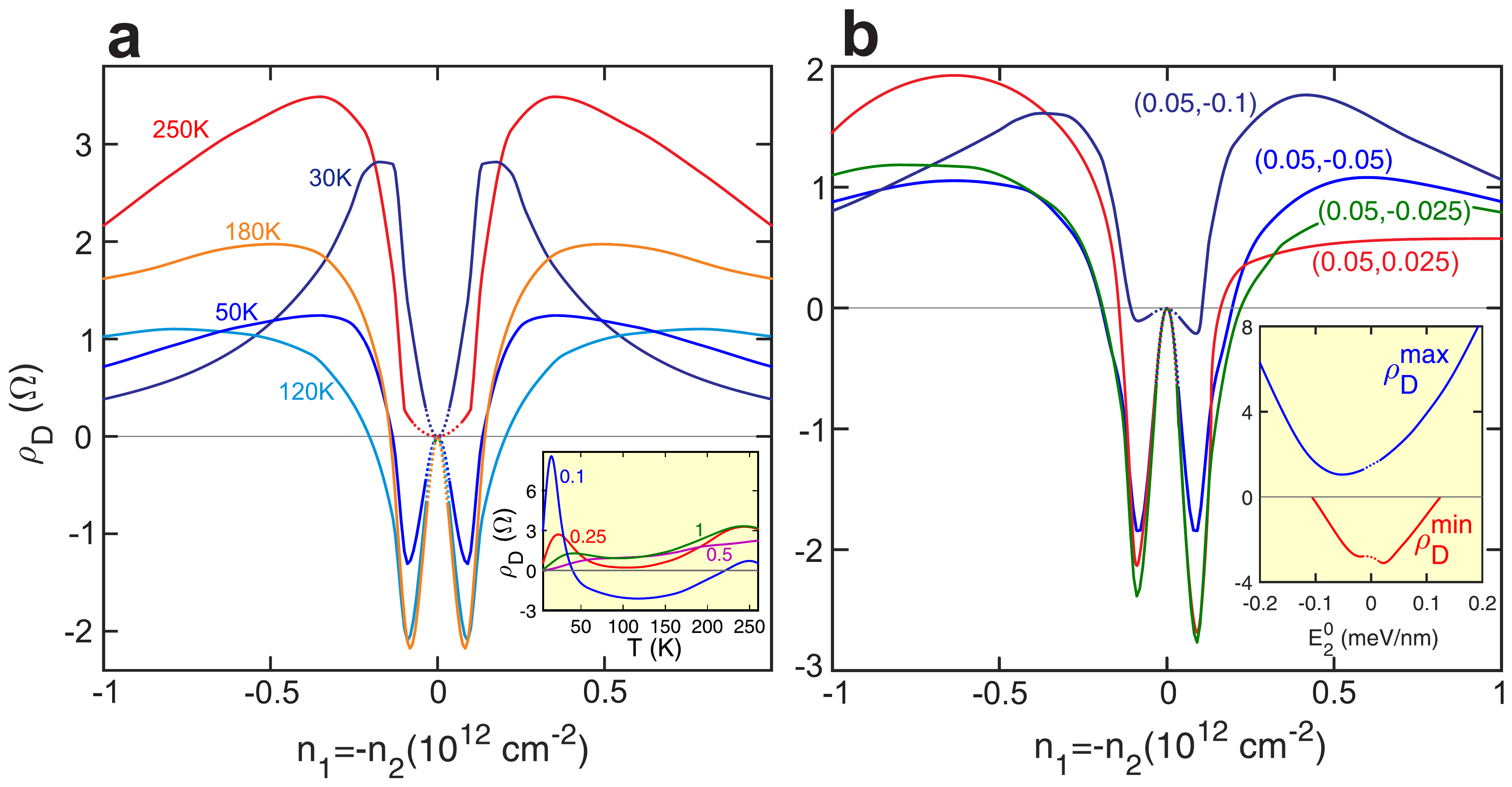}
\caption{
(a) Evolution with $T$ of the sign reversal and peaks of $\rho_D$.
For  $50\leq T \leq 180$ K the bilayers are almost intrinsic at low densities (see Fig. \ref{populations}(c))
resulting in a negative drag peak, while at higher densities the band populations are asymmetric (Fig. \ref{populations}(b)), making $\rho_D$ everywhere positive.
Inset shows the $T$ dependence of $\rho_D$. Curve labels are densities in $10^{12}$ cm$^{-2}$.
(b) Sensitivity of $\rho_D$ to the transverse electric fields $E_i$.
Curve labels are the fields at the DNP, ($E^0_1$, $E^0_2$) in meV.nm$^{-1}$. $T=70$ K.
Inset shows the magnitudes, in
$\Omega$, of the negative  and positive peaks  ($\rho_D^{\mathrm{min}}$,
$\rho_D^{\mathrm{max}}$)  as functions of $E^0_2$  for fixed $E^0_1=0.05$ meV.nm$^{-1}$.
}
    \label{e-hDrag}
  \end{figure}

Figure \ref{e-hDrag}(a) shows the  evolution of $\rho_D$  for different $T$
for barrier thickness $d=5.2$ nm and
$E^0_1=-E^0_2=0.05$ V.nm$^{-1}$.
Note for $ T \lesssim 30$ K we predict $\rho_D$ will be everywhere positive, since thermal excitations
are small.  When $T$ increases, the thermal excitations
make the bilayers  nearly intrinsic at lower densities (see Fig. \ref{populations}(c)), while
at higher densities, the $\mu_i$  have moved to the CB and VB,
resulting in asymmetric band populations and a positive $\rho_D$ (see Fig. \ref{populations}(b)).
For $T \gtrsim 250$ K,  the reduction in $m^\ast$
away from the DNP drives the $\mu_i$ rapidly to the CB or VB,
again resulting in asymmetric band populations and a positive $\rho_D$.
The inset shows the $T$ dependence of $\rho_D$  at  fixed densities. Consistent with
the experimental data in
the inset of Fig.\ 2(c) in  Ref.\ \citenum{li_negativeDrag}, at low densities $\rho_D$ changes sign:
from positive at small $T$ to negative at intermediate $T$,
and  back  to positive as $T$ approaches room temperature.
At higher densities, the $\rho_D$ is positive for all $T$ because the $\mu_i$ are in the CB and VB,
making the interband contributions negligible.

Figure \ref{e-hDrag}(b) highlights the sensitivity of the drag
to the transverse electric fields $E_i$.
 $\rho_D$ is shown at $T=70$ K for different combinations of  $E^0_i$.
The
barrier thickness
$d=5.2$ nm.
The
maximum and minimum $\rho_D$ both increase with $E^0_i$.
When the $E_i$ (and hence the bandgaps) are sufficiently large, $k_BT < \mu_i+E_i^g$,
and $\rho_D$
is positive for all densities, since thermal excitations are negligible.
When the  $E^0_1 \neq E^0_2$,  the $\rho_D$ is asymmetric upon
interchanging the e- and h-doped sheets.
The inset shows  the amplitudes of the positive and negative peaks
for fixed $E^0_1$ as a function of  $E^0_2$.
The  drag maximum  $\rho_D^{\mathrm{max}}$
and minimum  $\rho_D^{\mathrm{min}}$ both increase with $E_g^{i=2}$.
At very large $E^0_2$, the $\rho_D^{\mathrm{max}}$ saturates. A
one-band analysis is then sufficient.
The  drag becomes entirely positive when $E^0_2\gtrsim 0.1$ meV.nm$^{-1}$.
Thus, while the change in  sign of $\rho_D$ is relatively robust,
when the asymmetry of the two bandgaps becomes too pronounced,
the sign change is eventually lost.
By tuning the initial band gaps in
the
graphene bilayers $E_i^0$ with respect to the thermal excitation energy,
the sign of the drag at a fixed density and $T$ can be reversed.

No sign reversal of drag has been observed for e-h graphene monolayers or GaAs double quantum wells.
For graphene monolayers away from the DNP,
the $\mu_i$  move  rapidly
to the CB and VB,
and so
negative drag peaks  would only occur so close to the DNP that they would be undetectable.
For GaAs the large bandgap of $\sim 1.5$ eV means
$k_BT\ll E_g^i$
always,
so the $\mu_i$  remain
near
the CB and VB.  Then
the  $\mathfrak{Im}\chi_1(\boldsymbol{q},\omega)$
and  $\mathfrak{Im}\chi_2(\boldsymbol{q},\omega)$ have opposite signs, leading  to positive $\rho_D$.


The main limitation of our approach is
that
our calculations are made within RPA in the clean limit.
Our agreement with the experimental peak structures of the
$\rho_D$ indicates that the
effects of disorder
and
correlations
can have no major impact on the position of the $\rho_D$ peaks at fixed $T$.
Possible effects of superfluidity\cite{perali_dbg,zarenia_enhancement}
on the drag would occur at much lower $T$ than we consider.

This work reveals a new mechanism of drag in coupled multiband 2D sheets with small bandgaps,
and our multiband theory predicts negative drag not only in DBLG but also in related systems.
Our theory shows that for small bandgap systems,
multiband effects have a dramatic effect on the drag even for a Fermi liquid.
The reversals of sign in $\rho_D$ observed in Refs.\ \citenum{lee_giantDrag,li_negativeDrag} as high as
$T\sim 200$ K\cite{li_negativeDrag} suggest
that correlation effects, e-h puddles, or superfluidity
are not the primary mechanisms for the observed anomalous drag.
The structure in the drag
from  multiband effects needs to be fully mapped out.  Correlations, e-h puddles, and superfluidity should be studied
in the presence of multiband effects
in BLG, in trilayer graphene, and in other new 2D materials with small or zero bandgaps.
Our theory also predicts that hybrid systems containing one small and one large bandgap material
could also exhibit changes of sign in $\rho_D$, as
recently
observed in a BLG-GaAs hybrid system.\cite{Simonet2017}

\noindent {\it Acknowledgements.}  We are grateful to Cory Dean, Emanuel Tutuc, and their research groups
for discussing details of their experiments with us.
This work was partially supported by the Flemish Science
Foundation (FWO-Vl) and the Methusalem program of the Flemish government.
A.R.H. acknowledges support from the ARC Centre of Excellence for Future Low Energy Electronics Technologies.
D.N. acknowledges support from the University of Camerino FAR project CESEMN.


%

\end{document}